\documentclass[aps,prc,twocolumn,superscriptaddress,showpacs]{revtex4}
\usepackage{graphicx}

\newcommand{\beq}{\begin{equation}}
\newcommand{\eeq}{\end{equation}}

\begin{document}

\title{Pygmy dipole resonance in $^{208}$Pb}

\newcommand{\RCNP}{Research Center for Nuclear Physics, Osaka University, Ibaraki, Osaka 567-0047, Japan}
\newcommand{\Wit}{School of Physics, University of the Witwatersrand, Johannesburg 2050, South Africa}
\newcommand{\Kyushu}{Department of Physics, Kyushu University, Fukuoka 812-8581, Japan}
\newcommand{\iThemba}{iThemba LABS, Somerset West 7129, South Africa}
\newcommand{\Osaka}{Department of Physics, Osaka University, Toyonaka, Osaka 560-0043, Japan}
\newcommand{\CYRIC}{Cyclotron and Radioisotope Center, Tohoku University, Sendai,  980-8578, Japan}
\newcommand{\CNS}{Center for Nuclear Study, University of Tokyo, Bunkyo,  Tokyo 113-0033, Japan}
\newcommand{\TUDarmstadt}{Institut f\"ur Kernphysik, Technische Universit\"{a}t Darmstadt, D-64289 Darmstadt, Germany}
\newcommand{\Valencia}{Instituto de Fisica Corpuscular, CSIC-Universidad de Valencia, E-46071 Valencia, Spain}
\newcommand{\MSU}{NSCL, Michigan State Univ., MI 48824, USA}
\newcommand{\Kyoto}{Department of Physics, Kyoto University, Kyoto 606-8502, Japan}
\newcommand{\Niigata}{Department of Physics, Niigata University, Niigata 950-2102, Japan}
\newcommand{\RIKEN}{RIKEN Nishina Center, Wako, Saitama 351-0198, Japan}
\newcommand{\HIMAC}{National Institute of Radiological Sciences, Chiba 263-8555, Japan}
\newcommand{\ECT}{ECT*, Villa Tambosi, I-38123, Villazzano (Trento), Italy}
\newcommand{\KVI}{Kernfysisch Versneller Instituut, University of Groningen, Zernikelaan 25, NL-9747 AA Groningen, The Netherlands}
\newcommand{\TexasAM}{Department of Physics and Astronomy, Texas A\&M University-Commerce, Commerce, Texas 75429, USA}
\newcommand{\GSI}{GSI Helmholtzzentrum f\"{u}r Schwerionenforschung, D-64291 Darmstadt, Germany}
\newcommand{\EMMI}{ExtreMe Matter Institute EMMI and Research Division, GSI Helmholtzzentrum f\"{u}r Schwerionenforschung, D-64291 Darmstadt, Germany}

\author{I.~Poltoratska}\affiliation{\TUDarmstadt}
\author{P.~von~Neumann-Cosel}\email{vnc@ikp.tu-darmstadt.de}\affiliation{\TUDarmstadt}
\author{A.~Tamii}\affiliation{\RCNP}
\author{T.~Adachi}\affiliation{\Osaka} \affiliation{\KVI}
\author{C.~A.~Bertulani}\affiliation{\TexasAM}
\author{J.~Carter}\affiliation{\Wit}
\author{M.~Dozono}\affiliation{\Kyushu}
\author{H.~Fujita}\affiliation{\RCNP}
\author{K.~Fujita}\affiliation{\Kyushu}
\author{Y.~Fujita}\affiliation{\Osaka}
\author{K.~Hatanaka}\affiliation{\RCNP}
%\author{D.~Ishikawa}\affiliation{\RCNP}
\author{M.~Itoh}\affiliation{\CYRIC}
\author{T.~Kawabata}\affiliation{\Kyoto}
\author{Y.~Kalmykov}\affiliation{\TUDarmstadt}
\author{A.~M.~Krumbholz}\affiliation{\TUDarmstadt}
\author{E.~Litvinova}\affiliation{\EMMI}
\author{H.~Matsubara}\affiliation{\CNS}
\author{K.~Nakanishi}\affiliation{\CNS}
\author{R.~Neveling}\affiliation{\iThemba}
\author{H.~Okamura}\affiliation{\RCNP}
\author{H.~J.~Ong}\affiliation{\RCNP}
\author{B.~\"{Ozel-Tashenov}}\affiliation{\GSI}
\author{V.~Yu.~Ponomarev}\affiliation{\TUDarmstadt}
\author{A.~Richter}\affiliation{\TUDarmstadt}\affiliation{\ECT}
\author{B.~Rubio}\affiliation{\Valencia}
\author{H.~Sakaguchi}\affiliation{\RCNP}
\author{Y.~Sakemi}\affiliation{\CYRIC}
\author{Y.~Sasamoto}\affiliation{\CNS}
\author{Y.~Shimbara} \affiliation{\Osaka} \affiliation{\Niigata}
\author{Y.~Shimizu}\affiliation{\RIKEN}
\author{F.~D.~Smit}\affiliation{\iThemba}
\author{T.~Suzuki}\affiliation{\RCNP}
\author{Y.~Tameshige}\affiliation{\HIMAC}
\author{J.~Wambach}\affiliation{\TUDarmstadt}
%\author{R.~Yamada}\affiliation{\Niigata}
\author{M.~Yosoi}\affiliation{\RCNP}
\author{J.~Zenihiro}\affiliation{\RCNP}

\date{\today}

\begin{abstract}

Scattering of protons of several hundred MeV
is a promising new spectroscopic tool for the study of electric dipole strength in nuclei.
A case study of  $^{208}$Pb shows that at very forward
angles $J^\pi = 1^{-}$ states are strongly populated via Coulomb excitation.
A separation from nuclear excitation of other modes is achieved by a multipole decomposition analysis of the experimental cross sections based on theoretical angular distributions calculated within the quasiparticle-phonon model.
The B($E1$) transition strength distribution is extracted for excitation energies up to 9 MeV, i.e., in the region of the so-called pygmy dipole resonance (PDR).
The Coulomb-nuclear interference shows sensitivity to the underlying structure of the $E1$ transitions, which allows for the first time an experimental extraction of the electromagnetic transition strength and the energy centroid of the PDR.

\end{abstract}


\maketitle

% Introduction
The electric dipole ($E1$) strength in nuclei is dominated by the isovector giant dipole resonance (GDR), originating from the collective motion of neutrons against protons in the nucleus and located well above the particle emission threshold. The GDR provides basic insight into the isovector properties of the nuclear force and thus was intensively investigated both experimentally and theoretically \cite{ber75,har01}. However, at present the interest is more focused on low-lying dipole strength, well below the GDR energies, referred to as pygmy dipole resonance. It appears in nuclei with neutron excess and might be pictured macroscopically to result from oscillations of these excess neutrons against an inert core with $N \simeq Z$ (see, e.g., Ref.~\cite{paa07} and refs.\ therein). The PDR is predicted in all microscopic calculations based on the random-phase approximation (RPA), but the predicted central energy and strength often differ considerably, in particular between those based on non-relativistic and relativistic mean-field approaches (see e.g.\ Ref.~\cite{vnc10} for an example in the stable tin isotopes).

By its nature, the PDR may shed light onto the formation of neutron skins in nuclei \cite{pie06,kli07,tso08} and, because of the strong correlation in the RPA models, in turn on the symmetry energy \cite{kli07,car10,pie11}. Constraints on the magnitude and density dependence of the symmetry energy are important ingredients for the modeling of neutron stars. There is a clear correlation between the total electric dipole polarizability and the neutron skin \cite{rei10,pie11}. However, it has been argued that the PDR alone carries independent information \cite{kli07,pie11}.

The properties of the PDR in stable nuclei have been studied extensively for different neutron and proton shell closures with the $(\gamma,\gamma')$ reaction (e.g.\ Ref.~\cite{sav08} and references therein). Crucial data in exotic neutron-rich nuclei, where the PDR should be enhanced, are still scarce \cite{kli07,adr05,wie09} and suffer from large systematic uncertainties. The heaviest stable doubly magic nucleus $^{208}$Pb has always been a benchmark and the PDR has been studied in various recent $(\gamma,\gamma')$ experiments \cite{rye02,end03,shi08,sch10}. Theoretically, closed-shell nuclei permit the inclusion of complex degrees of freedom beyond the mean-field level in the microscopic calculations \cite{rye02,lit10}.

While the results of Refs.~\cite{rye02,end03,shi08,sch10} agree quite well, the $(\gamma,\gamma')$ reaction in general suffers from two problems: the experimental quantity measured is the ground-state gamma decay width times the ground-state branching ratio. The latter is usually not known and assumed to be 100\% in most analysis. However, statistical model calculations indicate potentially large correction factors (although not for the case of $^{208}$Pb) modifying the resulting PDR strength distributions considerably \cite{rus08}. Furthermore, the dominance of particle over $\gamma$ decay suppresses the experimental signal above threshold. Another, more general problem is the experimental separation of the PDR from the GDR. While theoretical transition densities provide a signature in the model calculations \cite{vre01}, the experimentally determined reduced B($E1$) strengths do not allow such a distinction. A possible experimental approach to the distinction of PDR and GDR is the use of isoscalar probes, and a pioneering $(\alpha,\alpha'\gamma)$ experiment has been performed \cite{pol92} demonstrating a large exhaustion of the isoscalar energy-weighted sum rule by low-energy $E1$ transitions in $^{208}$Pb.

Recently, a new experimental technique utilizing polarized proton scattering at and close to $0^\circ$ to measure the complete $E1$ strength in nuclei has been developed \cite{tam09}. It allows, in particular, a consistent extraction of the $E1$ transition strengths below and above the neutron threshold. At small momentum transfers and incident proton energies of several hundred MeV the cross sections are dominated by isovector spinflip-$M1$ transitions (the analog of the Gamow-Teller mode) \cite{hey10} and by Coulomb excitation of non-spinflip $E1$ transitions. A separation of these two contributions can be achieved either by a multipole decomposition analysis (MDA) of the angular distributions or by the analysis of polarization transfer observables. In Ref.~\cite{tam11} excellent agreement of the two methods was demonstrated for the case of $^{208}$Pb. Here we present our results for the $E1$ strength distribution at energies below the GDR based on the MDA. We also demonstrate that because of the interference of Coulomb and nuclear interaction, the angular distributions do show sensitivity to the underlying structure allowing for an experimental separation of PDR and GDR contributions.

\begin{figure}[hbt]
\includegraphics[width=8.6cm]{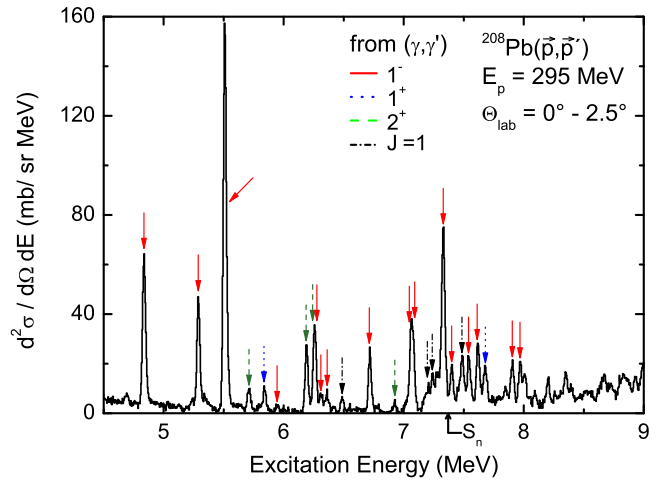}
\caption{\label{fig:spec}
(Color online) Low-energy part of the spectrum of the $^{208}$Pb($p,p'$) reaction at
E$_p$ = 295~MeV and $\Theta_{lab} = 0^\circ - 2.5^\circ$. The arrows indicate transitions also observed in $^{208}$Pb($\gamma$,$\gamma^\prime$) experiments~\cite{rye02,end03,shi08,sch10}. }
%Solid (red): $E1$ transitions. Dotted (blue): $M1$ transitions. Green %(dashed): $E2$ transitions.  Black (dashed-dotted): Dipole transitions %with unknown parity.}
\end{figure}

The $^{208}$Pb($\vec{p},\vec{p}'$) experiment was performed at the RING cyclotron facility
of the Research Center for Nuclear Physics (RCNP), Osaka University. A description of the experimental technique can be found in Ref.~\cite{tam09} and details of the present experiment and the polarization transfer measurements in Ref.~\cite{pol11}. A proton beam of 295 MeV with intensities $2-10$~nA and an average polarization of $P_0 \simeq 0.7$ bombarded an isotopically enriched $^{208}$Pb foil with an areal density of 5.2 mg/cm$^2$. Data were taken with the Grand Raiden spectrometer \cite{fuj99} in an angular range $0^\circ - 2.5^\circ$ and for excitation energies $E_x \simeq 4 - 22$ MeV.
%For the measurements of the polarization transfer coefficients $D_{SS'}$ and $D_{LL'}$ (for the definition see Ref.~\cite{ohl79}), sideways ($S$) and longitudinally ($L$) polarized proton beams, respectively, were utilized.
Additional data with unpolarized protons were taken at angles up to $10^\circ$. Employing dispersion matching techniques, an energy resolution $\Delta E \simeq 25$ keV (full width at half maximum) could be achieved. A spectrum of the $^{208}$Pb($p,p'$) reaction with the spectrometer set at $0^\circ$ is shown in Fig.~\ref{fig:spec} for the excitation energy range $4.5-9$~MeV, where the PDR is expected to lie. The arrows indicate the excitation energy of excited states identified in the ($\gamma$,$\gamma^\prime$) experiments \cite{rye02,end03,shi08,sch10}. Essentially, all prominent dipole transitions observed in the latter experiments are also excited in the present measurements.

\begin{figure}[tbh]
\includegraphics[width=8.6cm]{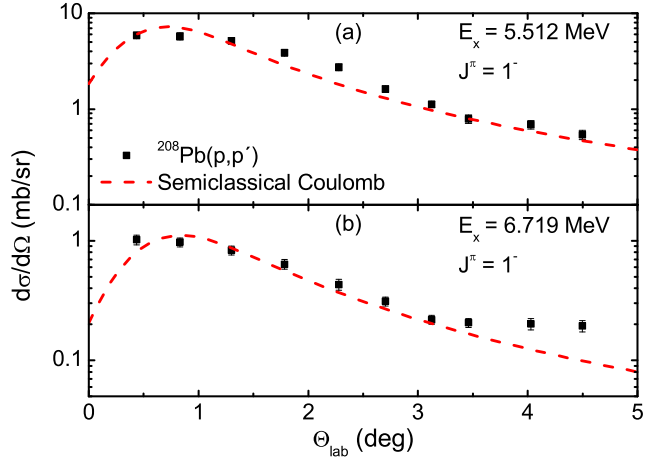}
\caption{\label{fig:Expvscoulomb}(Color online) Angular distributions of the prominently excited 1$^-$ states at $E_x=5.512$ MeV and 6.719 MeV in the $^{208}$Pb$(p,p')$ reaction at $E_p = 295$ MeV. The dashed lines are predictions of Coulomb excitation cross sections based on the semiclassical approach \cite{ber85}.}
\end{figure}

Excitation of $1^-$ states is possible through nuclear and Coulomb interaction, and both contributions add coherently to the cross sections. To verify the assumption of a predominant Coulomb excitation at angles close to $0^\circ$, predictions for the angular distributions in $(p,p'$) scattering were calculated based on a semiclassical model \cite{ber85}.
As examples, results for the prominent transitions to 1$^-$ states at $E_x =5.512$~MeV and 6.720~MeV are shown in Fig.~\ref{fig:Expvscoulomb}. Because of the finite angular resolution of the Grand Raiden spectrometer, the calculated cross section angular distributions were convoluted with Gaussian functions with widths corresponding to the vertical and horizontal angular opening of the detector system. The shape of the experimental angular distributions is well described and their absolute magnitudes can be reproduced when the calculations are normalized to the average B($E1$) strengths deduced from the ($\gamma$,$\gamma^\prime$) experiments \cite{rye02,end03,shi08,sch10}. The remaining deviations at angles larger than $2^\circ$ are attributed to effects of Coulomb-nuclear interference and in case of the transition to the state at 6.720 MeV to contributions from unresolved transitions with higher multipolarities.

In order to determine such contributions and to enable a separation from the spin-$M1$ resonance known to set in at $E_x > 7$ MeV \cite{hey10,las88}, a MDA was performed. The method, based on model predictions of the angular distribution shapes, is commonly used in the analysis of complex spectra from hadronic reactions, e.g.\ for an extraction of B(GT) strengths in charge-exchange reactions \cite{wak97} or isoscalar giant resonance strength distributions from inelastic $\alpha$-particle scattering \cite{li09}, and also for inelastic electron scattering form factors of nuclei \cite{str00}.
Theoretical proton scattering cross sections were calculated using the code DWBA07 \cite{dwba07} with RPA amplitudes and single-particle wave functions from the quasiparticle-phonon model (QPM) \cite{rye02}. The $t$-matrix parametrization of Love and Franey \cite{lov81} at 325~MeV was used as effective projectile-target interaction.
For each discrete transition ($E_x < 7$~MeV) or excitation energy bin ($E_x \geq 7$~MeV, cf.\ Tab.~\ref{tab:lowtransitions}), the experimentally obtained angular distributions were
fitted by means of the least-square method to a sum of the
calculated angular distributions weighted with coefficients $a^{E/M \lambda}$ (with the condition $a^{E/M \lambda} \geq 0$).

\begin{figure}[tbh]
\includegraphics[width=8.6cm]{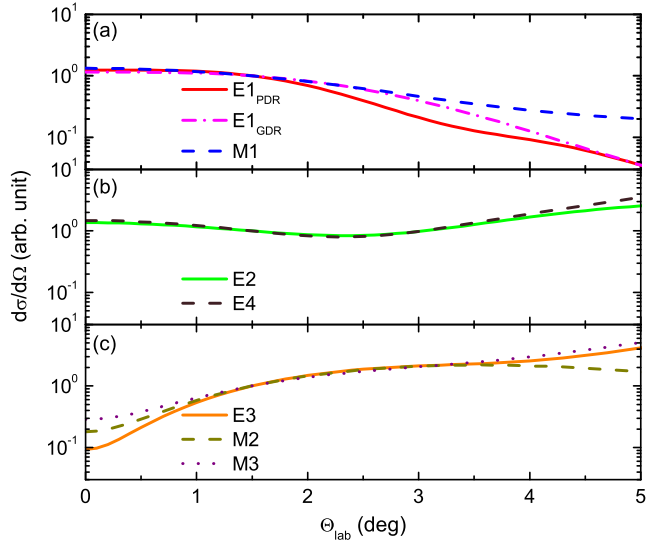}
\caption{\label{fig:theory}
(Color online) Comparison of theoretical angular distribution shapes used in the MDA: (a) $M1$ and representative $E1$ transitions to the PDR and GDR, (b) $E2$ and $E4$, (c) $E3$, $M2$ and $M3$. All curves are normalized at $\Theta_{\rm lab} = 1.5^\circ$.}
\end{figure}

Some approximations were necessary to make the MDA tractable:  Experimental data, although available up to $10^\circ$, were restricted to scattering angles $\Theta_{lab}\leq 4^\circ$ because of the increasing complexity of contributions from different multipoles at higher momentum transfers \cite{hof07}.
Isovector spin-$M1$ excitations were represented by a single characteristic curve, justified by the identical angular dependence of the cross section for all transitions of this type in the angular range considered. Furthermore, only $E1$ transitions with a strength larger than 0.01~e$^2$fm$^2$ were taken into account.
Figure \ref{fig:theory}(a) compares the shape of isovector spin-$M1$ transitions with representative examples of $E1$ transitions to states of the PDR and GDR, respectively. Indeed, the latter can not only be distinguished from the $M1$ case but also between each other.
All other contributions to the cross sections were substituted by angular distributions of either $E2$ or $E3$ transitions, whose shapes were taken to be that of the most collective transition of each type. Other multipolarities of potential relevance like $M2$, $M3$ or $E4$ exhibit very similar angular distribution shapes to either $E2$ or $E3$ as shown in Figs.~\ref{fig:theory}(b) and (c). Since isoscalar monopole transitions are only weakly excited in proton scattering and the giant monopole resonance is located at higher excitation energies than the region of interest, possible  contributions from $E0$ transitions were neglected.

\begin{figure}[tbh]
\includegraphics[width=8.6cm]{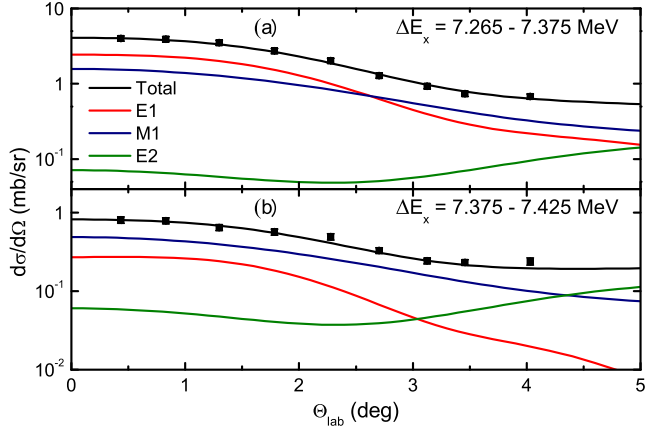}
\caption{\label{fig:mda}
(Color online) Examples of the MDA fits for two adjacent energy bins in the energy region of overlapping levels.}
\end{figure}

The final coefficients were obtained by computing the MDA for all possible combinations of $E1$, $M1$ and $E2$ (or $E3$) transitions and taking the $\chi^2$-weighted average of all individual $a^{E/M \lambda}$ values.
%
%For known electric dipole excitations in the energy region below 7.1~MeV the fitting procedure was simplified to include only theoretical transition amplitudes to the low-lying $1^-$ states. Only one $M1$ transition appears in the QPM calculations at low energies, which was used to describe the transition the 1$^+$ state at $E_x = 5.842$~MeV, known to have isoscalar character \cite{mue85}.
Examples of fits for two adjacent energy bins in the region of overlapping levels [see Figs.~\ref{fig:mda}(a) and (b)] demonstrate the sensitivity of the MDA to distinguish $E1$ and $M1$ contributions to the cross sections with a dominance of $E1$ in (a) and of $M1$ in (b), respectively.

Table~\ref{tab:lowtransitions} summarizes the MDA results for excitation energies $E_x = 4.8 - 9$~MeV. The partial $E1$ and $M1$ cross sections listed are integrated over a scattering angle range $0^\circ -0.94^\circ$, where nuclear contributions to the $E1$ cross sections are on the level of 1\% only. Thus, B($E1$) transition strengths can be directly derived.
\begin{table}[tbh]
\caption{$E1$ and spin-$M1$ cross sections in the $^{208}$Pb($p$,$p^\prime$) reaction for excitation energies $E_x = 4.8 - 9$ MeV integrated over scattering angles $0^\circ -0.94^\circ$ and derived B$(E1)$ strengths.}
\label{tab:lowtransitions}
\begin{center}
\begin{tabular}{l l l l l}
\hline\hline
$E_x$ &  $\sigma_{E1}$ & $\sigma_{M1}$ & $\sigma_{\mbox{total}}$ & B(E1) \\
(MeV)  & (mb) & (mb) & (mb) & (e$^2$fm$^2$) \\
\hline
 4.8420(22)  & 2.21(63) &  & 2.21(63) & 0.118(17)\\
 5.2949(22)  & 1.63(34) &  & 1.65(34)& 0.112(8)\\
 5.5128(11)  & 5.71(30) &  & 5.76(33)& 0.397(21)\\
 5.8417(50)  &  & 0.35(1) & 0.43(2) &           \\
 5.9463(59)  & 0.16(1) &  & 0.18(1)& 0.013(1)\\
 6.2642(26)  & 0.62(8) &  & 1.07(5) & 0.057(17)\\
 6.3131(59)  & 0.32(2) &  & 0.38(2)& 0.032(2)\\
 6.3585(65)  & 0.21(3) &  & 0.36(4)& 0.020(3)\\
 6.4835(49)  & 0.15(2) &  & 0.30(2)& 0.015(2)\\
 6.7184(26)  & 0.88(6) &  & 0.94(2) & 0.095(6)\\
7.005 - 7.135  & 2.06(3) & 0.22(1)& 2.305(16)   & 0.206(14)\\
7.135 - 7.225  & 0.48(10) & 0.28(8)& 0.776(3)  & 0.015(2)\\
7.225 - 7.265  & 0.41(9) & 0.25(7)&  0.681(2)  & 0.028(4)\\
7.265 - 7.375  & 2.47(39) & 1.52(30)& 4.016(150) & 0.254(23)\\
7.375 - 7.425  & 0.24(4) & 0.57(6)&  0.815(4)  & 0.021(3)\\
7.425 - 7.515  & 0.71(13) & 0.95(15)& 1.682(7) & 0.053(12)\\
7.515 - 7.585  & 0.72(15) & 0.63(14)& 1.362(20) & 0.061(13)\\
7.590 - 7.650  & 0.83(21) & 0.39(15)& 1.243(27)  & 0.109(25)\\
7.655 - 7.725  & 0.87(5) & 0.14(2)& 1.056(5)   & 0.104(6)\\
7.730 - 7.860  & 0.68(11) & 0.30(8)& 1.002(5)  & 0.072(18)\\
7.865 - 7.935  & 0.95(2) & 0.09(1)& 1.079(4)   & 0.120(18)\\
7.935 - 8.035  & 1.23(15) & 0.26(7)& 1.541(18)  & 0.167(18)\\
8.040 - 8.160  & 0.51(10) & 0.27(7)& 0.801(9)  & 0.055(13)\\
8.160 - 8.230  & 0.36(8) & 0.25(7)& 0.638(3)   & 0.052(12)\\
8.230 - 8.430  & 1.65(5) & 0.20(2)& 1.886(7)   & 0.242(15)\\
8.430 - 8.590  & 1.05(9) & 0.27(4)&  1.348(3)  & 0.145(20)\\
8.595 - 8.745  & 1.24(12) & 0.34(6)& 1.609(4)  & 0.191(25)\\
8.750 - 8.910  & 1.60(6) & 0.19(2)&  1.829(8)  & 0.277(23)\\
8.910 - 9.000  & 1.24(2) & 0.06(0)&  1.327(3)  & 0.215(24)\\
\hline\hline
\end{tabular}
\end{center}
\end{table}

\begin{figure}[tbh!]
\includegraphics[width=8.6cm]{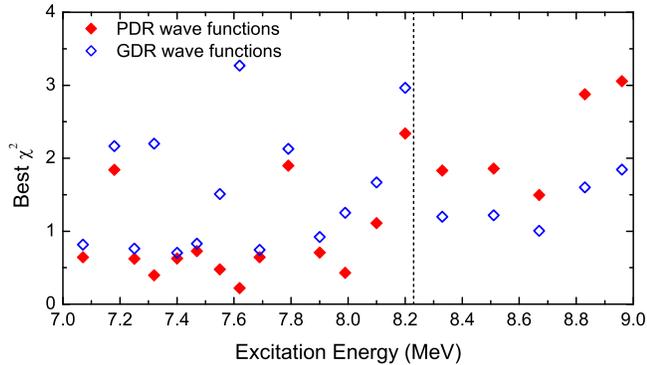}
\caption{\label{fig:bestchi}
(Color online) Best $\chi^2$ values in the MDA using either PDR- or GDR-type angular distributions for excitation energies $E_x = 7 - 9$ MeV and bins defined in Tab.~\ref{tab:lowtransitions}.}
\end{figure}

As discussed in Ref.~\cite{rye02}, the QPM calculations used for the present MDA analysis can distinguish between $E1$ transitions to the PDR and GDR by an analysis of the theoretical transition densities. The examples in Fig.~\ref{fig:theory}(a) demonstrate that the resulting $(p,p')$ angular distributions also show significant differences between excitation of these two modes. Thus, the MDA can provide information on the dominant structure of the $E1$ transitions. In Fig.~\ref{fig:bestchi}, the best $\chi^2$ values of  least-square fits using either PDR  (blue diamonds) or GDR (red diamonds) transition amplitudes are plotted as a function of the excitation energy. For $E_x \leq 8.23$ MeV,  $\chi^2$ results  assuming a PDR structure of the $E1$ excitations are consistently superior to fits with GDR-type angular distributions, while the reverse is observed for higher excitation energies. This finding allows to experimentally extract the properties of the PDR in $^{208}$Pb, viz.\ a centroid energy $E_{\ rm c} = 6.82(2)$ MeV and a strength $\sum{\rm B}(E1) = 2.18(7)$ e$^2$fm$^2$. The systematic uncertainty of the MDA approach is estimated to be at most 10\%  by constructing mixed PDR/GDR angular distributions with given amplitude ratios and studying the variation of $\chi^2$ in the fits.

\begin{figure}[!htb]
\includegraphics[width=8.6cm]{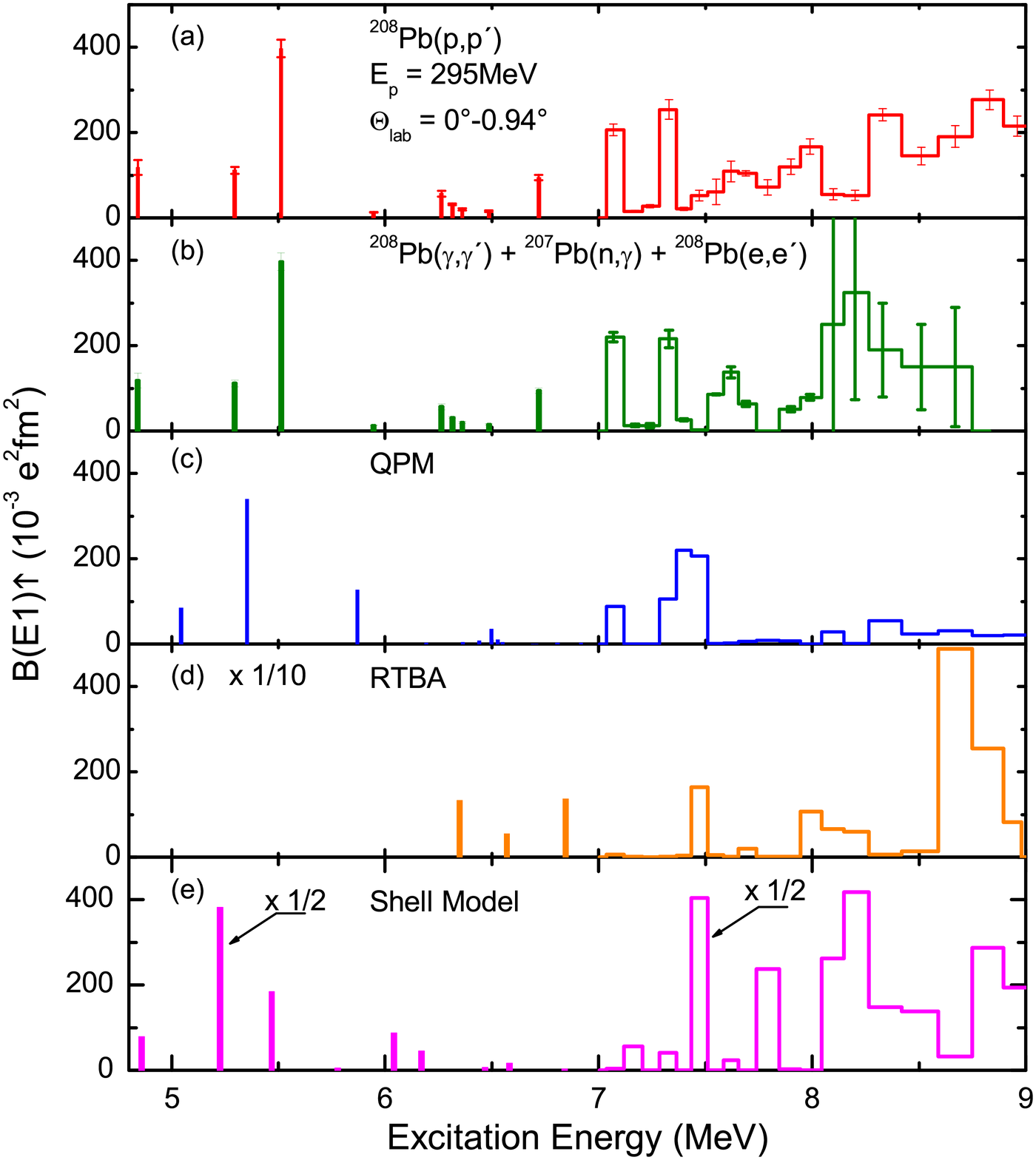}
\caption{\label{fig:BE1}
(Color online) E1 strength distributions in $^{208}$Pb between 4.8~MeV and 9~MeV from (a) the present experiment in comparison with (b) previous results and theoretical calculations within the (c) QPM,  (d) RTBA and (e) shell model. Note the scale reduction by a factor of 10 for the RTBA results.}
\end{figure}

Figure~\ref{fig:BE1} compares the electric dipole strength distribution up to $E_x = 9$ MeV extracted from the present experiment (a) with  previous results combining ($\gamma$,$\gamma^\prime$)   \cite{rye02,end03,shi08,sch10} and $^{207}$Pb(n,$\gamma)$ \cite{nndc} data up to 8 MeV with a $^{208}$Pb(e,e') measurement \cite{kue81} at higher excitation energies (b). The agreement is excellent up to the neutron separation energy ($S_n = 7.33$ MeV). Above threshold the present work shows additional, previously unobserved strength. This can most likely be attributed either to unknown neutron partial decay widths of the excited states. Above 8 MeV, $E1$ strength from the present work is consistent with the results of Ref.~\cite{kue81} within the systematic uncertainties of the latter.

Comparison with theoretical results from the quasiparticle-phonon model (QPM) \cite{rye02}, the relativistic time-blocking approximation (RTBA) \cite{lit07}, and a shell-model (SM) calculation  based on the interaction described in Ref.~\cite{bro00} is displayed in Figs.~\ref{fig:BE1}(c)-(e), respectively. The QPM and RTBA approaches have been discussed in detail in Ref.~\cite{tam11} and the SM in Ref.~\cite{sch10}. The QPM, including a model space up to 3-phonon states, provides a good description up to 7.5 MeV. At higher excitation energies the strength is clearly too small. The RTBA calculation so far includes beyond the 1 particle-1 hole ($1p1h$) states only configurations of the type $1p1h\otimes{\rm phonon}$, and thus the fragmentation is insufficient to describe the data (note the general reduction by a factor of 10 in  Fig.~\ref{fig:BE1}(d) to bring it on the scale of the other results). The total strength up to 9 MeV is more than a factor two too large. The SM shows a fair agreement for the total strength but the fragmentation is somewhat underpredicted.

It would be interesting to extract from the models centroid and strength of the PDR in $^{208}$Pb to be compared with the experimental result quoted above. However, we refrain from giving the corresponding values summed over the interval where the PDR is found experimentally. In the models the properties of the PDR are very sensitive to the mean-field description and predicted strength might thus be partially shifted to higher excitation energies. One should rather analyze the theoretical transition densities \cite{vre01,yuk12} to select those with a dominant PDR character.

To summarize, high-energy proton scattering at angles close to and including $0^\circ$ is used as a new experimental method to determine the B($E1$) strength distribution below the GDR in $^{208}$Pb. Combined with dispersion matching techniques, it provides a novel spectroscopic tool for the study of the electric dipole response in nuclei. The method also overcomes some limitations of other commonly used experimental techniques like $(\gamma,\gamma')$ or $(\gamma,n)$ reactions, which are sensitive to assumptions about decay branching ratios and typically limited to energy regions either below or above the neutron separation energy.

The MDA exhibits sensitivity to the structure of $E1$ transitions in $^{208}$Pb  through the difference of Coulomb-nuclear interference contributions to the cross sections of PDR and GDR excitations. This is an important finding because the strength distribution does not allow a distinction between the different modes. Thereby, the electromagnetic transition strength and the centroid energy of the PDR could be determined for the first time, providing an experimental benchmark for the strongly varying theoretical predictions (cf.\ Fig.~\ref{fig:BE1}). The method is not restricted to closed-shell nuclei but can be applied in vibrational and well deformed nuclei where microscopic calculations provide a good description of nuclear excitations. Possible approaches how to extract the PDR content from model calculations of the $E1$ response have been discussed e.g.\ in Refs.~\cite{co09,lan09}. Alternative information on the PDR structure may be obtained from isoscalar probes \cite{roc12,vre12} as demonstrated by studies of the $(\alpha,\alpha'\gamma)$ reaction \cite{pol92,end10}.

This work was supported by DFG (contracts SFB 634 and NE 679/3-1) and JSPS (Grant No.~14740154). B.~R.\ acknowledges support by the JSPS-CSIC collaboration program and E.~L.\ by the Helmholtz Alliance EMMI.

%\bibliography{pbprl}

\end{document}